\documentclass{elsart}
\usepackage[latin1]{inputenc}
\usepackage{amsmath}
\usepackage{amsfonts}
\usepackage{amssymb}
\usepackage{latexsym}

\newcommand{\uple}[1]{{\langle #1 \rangle}}
\newcommand{\R}{\mathbb{R}}
\newcommand{\N}{\mathbb{N}}
\newcommand{\Q}{\mathbb{Q}}

\newcommand{\A}{\mathcal{A}}
\newcommand{\B}{\mathcal{B}}
\newcommand{\C}{\mathcal{C}}
\newcommand{\D}{\mathcal{D}}

\newcommand{\F}{\mathcal{F}}

\newcommand{\M}{\mathcal{M}}
\renewcommand{\O}{\mathcal{O}}
\renewcommand{\P}{\mathcal{P}}
\renewcommand{\S}{\mathcal{S}}
\newcommand{\U}{\mathcal{U}}
\newcommand{\V}{\mathcal{V}}
\newcommand{\X}{\mathcal{X}}
\newcommand{\Y}{\mathcal{Y}}
\newcommand{\FI}{\varphi}

\newcommand{\vers}{\rightarrow}

\newcommand{\comp}[1]{{#1}^\mathcal{C}}
\newcommand{\dom}{\mbox{dom}}

\newcommand{\defin}[1]{\emph{\textbf{#1}}}
\newcommand{\seq}[1]{\overrightarrow{#1}}
\newcommand{\orcl}[1]{^{[#1]}}
\newcommand{\puce}{$\bullet$ }

\newcommand{\CXY}{\mathcal{C}(X,Y)}
\newcommand{\Rplus}{\overline{\R}^+}
\newcommand{\restr}[2]{{#1}_{|_{#2}}}

\newtheorem{lemma}{Lemma}[subsection]
\newtheorem{proposition}{Proposition}[subsection]
\newtheorem{theorem}{Theorem}[subsection]
\newtheorem{corollary}{Corollary}[subsection]
\newtheorem{definition}{Definition}[subsection]
\newtheorem{remark}{Remark}[subsection]

\newenvironment{proof}{\noindent\textbf{proof: }}{\hspace{\stretch{1}} $\square$}

\newenvironment{examples}{\noindent\textbf{examples:}\begin{enumerate}}{\end{enumerate}}

\journal{Information and Computation}
\begin{document}
\begin{frontmatter}
\title{Computability of probability measures and\\ Martin-L\"of randomness over metric spaces}

\author{Mathieu Hoyrup}
\address{LIENS, Ecole Normale Sup\'erieure, Paris. email: hoyrup@di.ens.fr}
\author{Crist\'obal Rojas}
\address{LIENS, Ecole Normale Sup\'erieure and CREA, Ecole Polytechnique,
Paris. email: rojas@di.ens.fr}
\thanks{partly supported by ANR Grant  05 2452 260 ox}


\date{}

\begin{abstract}
In this paper we investigate algorithmic randomness on more general
spaces than the Cantor space, namely computable metric spaces. To do this, we
first develop a unified framework allowing computations with probability
measures. We show that any computable metric space with a computable
probability measure is isomorphic to the Cantor space in a
computable and measure-theoretic sense. We show that any computable metric
space admits a universal uniform randomness test (without further assumption).
\end{abstract}


\end{frontmatter}

\section{Introduction}

The theory of algorithmic randomness begins with the definition of
individual random infinite sequence introduced in 1966 by Martin-L\"of
\cite{MLof66}.  Since then, many efforts have contributed to the development of
this theory which is now well established and intensively studied, yet
restricted to the Cantor space. In order to carry out an extension of this
theory to more general infinite objects as encountered in most mathematical models of physical random phenomena,
a necessary step is to understand what means for a probability measure on a general space to be computable (this is very simple expressed on the Cantor Space). Only then algorithmic randomness can be extended.

The problem of computability of (Borel) probability measures over more general spaces has been investigated by several authors:
by Edalat for compact spaces using domain-theory (\cite{Eda96}); by Weihrauch
for the unit interval (\cite{Wei99}) and by Schr\"oder for sequential
topological spaces (\cite{Sch07}) both using representations; and by G\'acs for
computable metric spaces (\cite{Gac05}). Probability measures can be seen
from different points of view and those works develop, each in its own
framework, the corresponding computability notions. Mainly, Borel probability
measures can be regarded as points of a metric space, as valuations on open
sets or as integration operators. We express the computability counterparts of these different views in a unified framework, and show
them to be equivalent.

Extensions of the algorithmic theory of randomness to general spaces have
previously been proposed: on effective topological spaces by Hertling and
Weihrauch (see \cite{HerWei98},\cite{HerWei03}) and on computable metric
spaces by G\'acs (see \cite{Gac05}), both of them generalizing the notion
of randomness tests and investigating the problem of the existence of a
universal test. In \cite{HerWei03}, to prove the existence of such a test, ad
hoc computability conditions on the measure are required, which {\it a
posteriori} turn out to be incompatible with the notion of computable
measure. The second one (\cite{Gac05}), carrying the extension of Levin's theory of
randomness, considers \emph{uniform tests} which are tests parametrized by
measures. A computability condition on the basis of ideal balls (namely, recognizable Boolean inclusions) is needed
to prove the existence of a universal uniform test.

In this article, working in computable metric
spaces with any probability measure, we
consider both uniform and non-uniform tests and prove the following points:

\begin{itemize}
\item uniformity and non-uniformity do not essentially differ,
\item the existence of a universal test is assured without any further
condition.
\end{itemize}

Another issue addressed in \cite{Gac05} is the characterization of
randomness in terms of Kolmogorov Complexity (a central result in Cantor Space). There, this characterization is proved to hold (for a compact computable metric space $X$ with a computable measure) under the assumption that there exists a computable
injective encoding of a full-measure subset of $X$ into binary sequences. In the real line for example, the base-two numeral system (or binary expansion) constitutes such encoding for the Lebesgue measure. This fact was already been (implicitly) used in the definition of random reals (reals with a random binary expansion, w.r.t the uniform measure).

We introduce, for computable metric spaces with a computable measure, a notion of binary representation generalizing the base-two numeral system of the reals, and
prove that:
\begin{itemize}
\item such a binary representation always exists,
\item a point is random if and only if it has a unique binary expansion, which
is random.
\end{itemize}

Moreover, our notion of binary representation allows to identify any computable probability space
with the Cantor space (in a computable-measure-theoretic sense). It provides a tool to directly transfer elements of
algorithmic randomness theory from the Cantor space to any computable
probability space. In particular, the characterization of randomness in
terms of Kolmogorov complexity, even in a non-compact space, is a direct
consequence of this.

The way we handle computability on continuous spaces is largely inspired by
representation theory. However, the main goal of that theory is to study,
in general topological spaces, the way computability notions depend on the
chosen representation. Since we focus only on Computable Metric Spaces (see \cite{Hem02} for instance) and \emph{Enumerative Lattices} (introduced in setion 2.2) we shall consider only one \emph{canonical} representation for each set, so we do not use representation theory in its general setting. 

Our study of measures and randomness, although restricted to computable
metric spaces, involves computability notions on various sets which do not
have natural metric structures. Fortunately, all these sets become enumerative lattices in a very natural way and the canonical representation provides in each case the right computability notions.

In section \ref{basic}, we develop a language intended to express computability
concepts, statements and proofs in a rigorous but still (we hope)
transparent way. The structure of computable metric space is then
recalled. In section \ref{section_es}, we introduce the notion of enumerative lattices and present two important examples to be
used in the paper. Section \ref{measures} is devoted to the detailed study of
computability on the set of probability measures. In section \ref{cps} we define
the notion of binary representation on any computable metric space with a
computable measure and show how to construct such a representation. In
section \ref{randomness} we apply all this machinery to algorithmic randomness.

\section{Basic definitions}\label{basic}

\subsection{Recursive functions.}

The starting point of recursion theory was the mathematization of the
intuitive notion of function computable by an \emph{effective procedure} or \emph{algorithm}. The different systems and computation models
formalizing \emph{mechanical} procedures on natural numbers or symbols have turned
out to coincide, and therefore have given rise to a robust mathematical
notion which grasps (this is Church-Turing thesis) what means for a
(partial) function $\FI:\N\to\N$ to be \emph{algorithmic},
and which can be made precise using any one of the numerous formalisms
proposed. Following the usual denomination, we call such a function a \defin{(partial)
  recursive function}. To show that a function $\varphi:\N\to\N$ is
recursive, we will exhibit an algorithm $\A$ which on input $n$ halts and outputs
$\varphi(n)$ when it is defined, runs forever otherwise.

In the same vein, a robust notion of (partial) recursive function
$F:\N^\N\to\N^\N$ can be characterized by different formal definitions:

\begin{description}
\item[Via domain theory] (see \cite{AbrJun94}). This approach takes the
  notion of recursive function as primitive, which avoids the definition of
  a new computation model. A partial function $F:\N^\N\to\N^\N$ is
recursive if there is a recursive function $F':\N^*\to\N^*$
which is monotone for the prefix ordering, such that for all $\sigma\in
\dom(F)$, $F(\sigma)$ is the infinite sequence obtained at the limit by
computing $F'$ on the finite prefixes of $\sigma$ (precisely, the Baire
space can be embedded into the set of finite and infinite sequences of
integers ordered by the prefix relation, which is an $\omega$-algebraic
domain).
\item[Via oracle Turing machines] (used by Ko and Friedman, see
  \cite{KoFri82}, \cite{Ko91}). An oracle Turing machine $\M\orcl{\sigma}$
  is a Turing machine which works with a sequence $\sigma\in\N^\N$ provided
  as oracle and is allowed to read elements $\sigma_n$ of the oracle
  sequence. On an input $n\in\N$, it may stop and output a natural number,
  interpreted as $F(\sigma)_n$.
\item[Via type-two Turing machines] (defined by Weihrauch, see
  \cite{Wei00}). Expressed differently, it is essentially the same
  computation model (it works on symbols instead of integers).
\end{description}

Again, to show that a function $F:\N^\N\to\N^\N$ is recursive, we will
exhibit an algorithm $\A$ which given $\sigma\in\N^\N$ as oracle and $n$ as input,
halts and outputs $F(\sigma)_n$. The algorithm together with $\sigma$ in the oracle is denoted $\A\orcl{\sigma}$.

A sequence $\sigma\in\N^\N$ is \defin{recursive} if the function $n\mapsto
\sigma_n$ is recursive. Given a family $(\sigma_i)_{i\in\N}$ of recursive
sequences, $\sigma_i$ is \defin{recursive uniformly in $\boldsymbol{i}$} if
the function $\uple{i,n}\mapsto \sigma_{i,n}$ is recursive, where $\uple{,}$ denotes some computable bijection between tuples and natural numbers.

\subsection{Representations and constructivity}
A representation on a set $X$ is a surjective (partial) function $\rho:\N^\N\to X$. Let
$X$ and $Y$ be sets with fixed representations $\rho_X$ and $\rho_Y$.

\begin{definition}[Constructivity notions]\label{computability_notions}$\phantom{}$

\begin{enumerate}
\item An element $x\in X$ is \defin{constructive} if there is a recursive
sequence $\sigma$ such that $\rho_X(\sigma)=x$. 

\item The elements of a sequence $(x_i)_{i\in\N}$ are \defin{uniformly
constructive} if there is a family $(\sigma_i)_i$ of uniformly recursive
sequences such that $\rho_X(\sigma_i)=x_i$ for all $i$.

\item A function $f:\subseteq X\to Y$ is \defin{constructive} on $D\subseteq X$ if there
exists a recursive function $F:\N^\N\to\N^\N$ such that the following
diagram commutes on $\rho^{-1}_X(D)$:
$$
\begin{array}{rcl}
\N^\N & \stackrel{F}{\longrightarrow} & \N^\N \\
\rho_X \downarrow & & \downarrow \rho_Y \\
X & \stackrel{f}{\longrightarrow} & Y
\end{array}\qquad
(\mbox{that is, $f\circ \rho_X=\rho_Y\circ F$ on $\rho^{-1}_X(D)$})
$$
\end{enumerate}
\end{definition}

We say that $y$ is \defin{$\boldsymbol{x}$-constructive} if there is a function $f:\subseteq X\to Y$ constructive on $\{x\}$ with $f(x)=y$. If $x$ is
constructive, $x$-constructivity and constructivity are equivalent. Note that two sequences of natural numbers can be merged into
a single one, so the product $X\times Y$ of two represented sets has a
canonical representation. In particular, it makes sense to speak about
$(x,y)$-constructive elements.

\subsection{Objects}

There is a canonical way of defining a representation on a set $X$ when 1)
some collection of \emph{elementary objects} of $X$ can be encoded into natural numbers and 2) an
element of $X$ can be described by a sequence of these elementary
objects. Once encoded into natural numbers, the elementary objects inherit
their finite character and may be output by algorithms. Let us make it precise:

\begin{definition}
A \defin{numbered set} $\O$ is a countable set together with a total surjection
$\nu_\O:\N \to \O$ called the \defin{numbering}. We write $o_n$ for $\nu(n)$.
\end{definition}

A numbered set $\O$ and a (partial) surjection $\delta:\O^N\to X$ induce
canonically a representation $\rho=\delta\circ \nu_\mathcal{O}$. At least
in this paper, all representations will be obtained in this way. A sequence
of finite objects which is mapped by $\delta$ to $x$ is called a \defin{description} of
$x$.

An algorithm may then be seen as outputting objects:

Given a numbered set $\O$, we say that an algorithm (plain or with oracle)
\defin{enumerates} a sequence of objects $(o_{n_i})_{i\in\N}$ if on input
$i$ it outputs $n_i$. Given a representation $(\O,\delta)$ on a set $X$, an
algorithm enumerating a description of $x\in X$ is said to \defin{describe} $x$.

An algorithm may also take objects as inputs, with a restriction:

\begin{definition}
An algorithm $\A$ is said to be \defin{extensional} on an element
$x\in X$ if for all $\sigma$ such that $\rho_X(\sigma)=x$, $\A\orcl{\sigma}$
describes the same element $y\in Y$.
\end{definition}

We then say that $\A$ $x$-describes $y$ or that $\A\orcl{x}$ describes $y$.

The constructivity notions of definition \ref{computability_notions} can
then be expressed using this language, which will be used throughout this paper.

\begin{enumerate}
\item An element $x\in X$ is constructive if there is an algorithm describing $x$,

\item The elements of a sequence $(x_i)_{i\in\N}$ are uniformly
constructive if there is an algorithm $\A$ such that $\A(\uple{i,.})$ describes $x_i$,

\item A function $f:\subseteq X\to Y$ is constructive on $D\subseteq X$ if there exists an algorithm
  which $x$-describes $f(x)$ for all $x\in D$.
\end{enumerate}

A $x$-constructive element $y$ may be $x$-described by an algorithm which
is extensional only on $x$, and thus induce a function which is
defined only at $x$.

\subsection{Computable Metric Spaces}

\begin{definition}A \defin{computable metric space} is a triple
  $\X=(X,d,\S)$, where:

\puce $(X,d)$ is a separable complete metric space (polish metric space),

\puce $\S=\{s_i:i\in\N\}$ is a countable dense subset of $X$,

\puce The real numbers $d(s_i,s_j)$ are all computable, uniformly in $\uple{i,j}$.
\end{definition}

The elements of $\S$ are called the \defin{ideal points}. The numbering
$\nu_\S$ defined by $\nu_\S(i):=s_i$ makes $\S$ a numbered set. Without loss of
generality, $\nu_\S$ can be supposed to be injective: as $d(s_i,s_j)>0$ can be semi-decided, $\nu_S$ can be effectively transformed into an injective numbering. Then a sequence of ideal points can be uniquely identified with
the sequence of their names.

The numbered sets $\S$ and $\Q_{>0}$ induce the numbered set of \defin{ideal balls}
$\B:=\{B(s_i,q_j):s_i \in \S, q_j \in\Q_{>0}\}$, the numbering being
$\nu_\B(\uple{i,j}):=B(s_i,q_j)$. We write $B_{\uple{i,j}}$ for
$\nu_\B(\uple{i,j})$. The closed ball $\{x\in X:d(s,x)\leq
r\}$ is denoted $\overline{B}(s,q)$ and may not coincide with the closure
of the open ball $B(s,q)$ (typically, if the space has disconnection).

We now recall some important examples of computable metric spaces:

\begin{examples}
 \item $\mathcal(\Sigma^{\N},d,\S)$ where $\Sigma$ is a
finite alphabet, $d(\omega,\omega'):=2^{-\min\{n\in\N:\omega_n\neq
  \omega'_n\}}$ and $\S:=\{w000\ldots: w\in\Sigma^*\}$ where $\Sigma^*$ is
the set of finite words on $\Sigma$ and $0$ is some fixed symbol from $\Sigma$, 

\item $(\R^n,d_{\R^n},\Q^n)$ with the euclidean metric and the standard
numbering of $\Q^n$,

\item The product $(X\times Y,d,\S_X\times \S_Y)$ of two computable metric spaces has a canonical computable metric space structure, with $d((x,y),(x',y'))=\max\{d_X(x,x'),d_Y(y,y')\}$.

\end{examples}

For further examples, like functions spaces $C[0,1]$ and $L^p$ for
computable $p\geq 1$ we refer to [Weihrauch]. A sequence $(x_n)_{n\in\N}$ of points is said to be a \defin{fast Cauchy} sequence, or simply a \defin{fast} sequence if $d(x_n,x_{n+1})< 2^{-n}$ for all $n$.

\begin{definition}
On a computable metric space $(X,d,\S)$, the canonical representation is the Cauchy representation $(\S,\delta_C)$ 
defined by $\delta_C(\seq{s})=x$ for all fast sequence $\seq{s}$ of ideal
points converging to $x$.
\end{definition}

Again, each set $X$ with a computable metric structure $(X,d,\S)$ will be implicitly represented using the
Cauchy representation. Then canonical constructivity notions
derive directly from definition \ref{computability_notions}. It is usual to
call a constructive element of $X$ a \defin{computable point}, and a
constructive function between computable metric space, a \defin{computable function}. Remark
that the computable real numbers are the computable points of the computable metric space $(\R,d,\Q)$.

The choice of this representation is justified by the classical result: every computable function between computable metric spaces is continuous (on its domain of computability).

\begin{proposition}
The distance $d:X\times X \to \R$ is a computable function.
\end{proposition}

\begin{proposition}\label{proposition_computable_point}For a point $x\in X$, the following statements are equivalent:

\puce $x$ is a computable point,

\puce all $d(x,s_i)$ are upper semi-computable uniformly in $i$,

\puce $d_x:=d(x,.):X\to\R$ is a computable function.
\end{proposition}

Several metrics and effectivisations of a single set are possible, and induce in general different computability notions: two computable metric structures $(s,\S)$ and $(d',\S')$ are said to be \emph{effectively equivalent} if $id:(X,d,\S)\to(X,d',\S')$ is a computable homeomorphism (with computable inverse). In this case, all computability notions are preserved replacing one structure by the other (see \cite{Hem02} for details).

\section{Enumerative Lattices}\label{section_es}
\subsection{Definition}

We introduce a simple structure using basic order theory, on which a
natural representation can be defined. The underlying ideas are those from domain theory, but the framework is lighter and (hence) less powerful. Actually, it is sufficient for the main purpose: proposition \ref{enum_es}. This will be applied in the last section on randomness.
\begin{definition}
An \defin{enumerative lattice} is a triple $(X,\leq,\P)$ where
$(X,\leq)$ is a complete lattice and $\P\subseteq X$ is a numbered set such that
every element $x$ of $X$ is the supremum of some subset of $\P$.
\end{definition}

We then define $\P_\downarrow(x):=\{p\in\P: p\leq x\}$ (note that $x=\sup \P_\downarrow(x)$). Any element of $X$ can be described by a sequence $\seq{p}$ of elements of $\P$. Note that the least element $\bot$ need not belong to $\P$: it can be described by the empty set, of which it is the supremum.

\begin{definition}\label{representation_es}
The canonical representation on an enumerative lattice $(X,\leq,\P)$ is the induced by
the partial surjection $\delta_{\leq}(\seq{p})=\sup \seq{p}$ (where the sequence $\seq{p}$ may be empty).
\end{definition}

From here and beyond, each set $X$ endowed with an enumerative structure
$(X,\leq,\P)$ will be implicitly represented using the canonical
representation. Hence, canonical constructivity notions derive directly
from definition \ref{computability_notions}. Let us focus on an example:
the identity function from $X$ to $X$ is computed by an algorithm outputting
exactly what is provided by the oracle. Hence, when the oracle is empty,
which describes $\bot$, the algorithm runs forever and outputs nothing,
which is a description of $\bot$.


\begin{examples}
\item $(\overline{\R},\leq,\Q)$ with $\overline{\R}=\R \cup \{-\infty,+\infty\}$: the constructive elements are the so-called \emph{lower semi-computable} real numbers,
\item $(2^\N,\subseteq,\{\mbox{finite sets}\})$: the constructive elements are the r.e sets from classical recursion theory,
\item $(\{\bot,\top\},\leq,\{\top\})$ with $\bot<\top$.
\end{examples}

We recall that a real number $x$ is \emph{computable} if both $x$ and $-x$ are
lower semi-computable.

Here is the main interest of enumerative lattices:

\begin{proposition}\label{enum_es}
Let $(X,\leq,\P)$ be an enumerative lattice. There is an enumeration $(x_i)_{i\in\N}$ of
all the constructive elements of $X$ such that $x_i$ is constructive
uniformly in $i$.
\end{proposition}

\begin{proof}
there is an enumeration $\FI$ of the r.e subsets of $\N$: for every r.e subset $E$ of $\N$, there is some $i$ such that $E=E_i:=\{\FI(\uple{i,n}):n\in\N\}$. Moreover, we can take $\FI$ such that whenever $E_i\neq\emptyset$ the function $\FI(\uple{i,.}):\N\to\N$ is total (this is a classical construction from recursion theory, see \cite{Rog87}). Then consider the associated algorithm $\A_\FI=\nu_\P\circ \FI$: for every constructive element $x$ there is some $i$ such that $\A_\FI(\uple{i,.}):\N\to \P$ enumerates $x$ ($\emptyset$ is an enumeration of $\bot$).
\end{proof}

\begin{remark}\label{scott}
Observe that on every enumerative lattice the Scott topology can be defined: a Scott open set $O$ is an upper subset ($x\in O, x\leq y \Rightarrow y\in O$) such that for each sequence $\seq{p}=(p_{n_i})_{i\in\N}$ such that $\sup \seq{p}\in O$, there is some $k$ such that $\sup\{p_{n_0},\ldots,p_{n_k}\}\in O$.

If $Y$ and $Z$ have enumerative lattice structures, a function $f:Y\to Z$ is said to be Scott-continuous if it is monotonic and commutes with suprema of increasing sequences (one can prove that $f$ is Scott-continuous if and only if it is continuous for the Scott topologies on $Y$ and $Z$) and is easy to see that a Scott-continuous function $f:Y\to Z$ such that all $f(\sup\{p_{n_1},\ldots,p_{n_k}\})$ are constructive uniformly in $\uple{n_1,\ldots,n_k}$, is in fact a constructive function.
\end{remark}

\subsection{Functions from a computable metric space to an enumerative lattice}\label{section_CXY}

Given a computable metric space $(X,d,\S)$ and an enumerative space $(Y,\leq,\P)$, we
define the numbered set $\mathcal{F}$ of \defin{step functions} from $X$ to $Y$:
$$
f_\uple{i,j}(x)=\left\{
\begin{array}{rl}
p_j & \mbox{ if $x\in B_i$} \\
\bot & \mbox{ otherwise}
\end{array}\right.
$$

We then define $\CXY$ as the closure of $\F$ under pointwise suprema, with the
pointwise ordering $\sqsubseteq$. We have directly:

\begin{proposition}
$(\CXY,\sqsubseteq,\mathcal{F})$ is an enumerative lattice.
\end{proposition}

\textbf{example:} the set $\Rplus=[0,+\infty)\cup\{+\infty\}$ has an enumerative lattice
  structure $(\Rplus,\leq,\Q^+)$, which induces the enumerative lattice $\C(X,\Rplus)$
of positive lower semi-continuous functions from $X$ to $\Rplus$. Its constructive
elements are the positive \emph{lower semi-computable} functions.\\

We now show that the constructive elements of $\CXY$ are exactly the constructive functions from $X$ to $Y$.

To each algorithm $\A$ we associate a constructive
element of $\CXY$, enumerating a sequence of step functions: enumerate all
$\uple{n,i_0,\ldots,i_k}$ with $d(s_{i_j},s_{i_{j+1}})<2^{-(j+1)}$ for all
$j<k$ (prefix of a super-fast sequence). Keep only those for which the
computation of $\A\orcl{i_0,\ldots,i_k,0,0,\ldots}(n)$ halts without trying to
read beyond $i_k$. For each one, the latter computation outputs some element
$p_l$: then output the step function $f_\uple{i,l}$ where
$B_i=B(s_{i_k},2^{-k})$. We denote by $f_\A$ the supremum of the enumerated sequence of step functions.

\begin{lemma}\label{lemma_extensional}
For all $x$ on which $\A$ is extensional, $f_\A(x)$ is the element of $Y$ described by $\A\orcl{x}$.
\end{lemma}

\begin{proof}
let $y$ be the element described by $\A\orcl{x}$.

For all $\uple{n,i_0,\ldots,i_k}$ for which some $f_\uple{i,j}$ is
enumerated with $x\in B_i$, there is a fast sequence $\seq{s}$ converging to $x$ starting with $s_{i_0},\ldots,s_{i_k}$, for which $\A\orcl{\seq{s}}(n)=p_j$. Then $y\geq p_j=f_\uple{i,j}(x)$. Hence $y\geq f_\A(x)$.

There is a super-fast sequence $\seq{s}$ converging to $x$: for all $n$, $\A\orcl{\seq{s}}(n)$ stops and outputs some $p_{j_n}$, so there is some $i_n$ with $x\in B_{i_n}$ such that $f_\uple{i_n,j_n}$ is enumerated. Hence, $y=\sup_n p_{j_n}=\sup f_\uple{i_n,j_n}(x)\leq f_\A(x)$. 
\end{proof}

\begin{proposition}\label{proposition_constructive_functions}
The constructive elements of $\CXY$ are exactly the (total) constructive
functions from $X$ to $Y$.
\end{proposition}

\begin{proof}
the supremum of a r.e subset $E$ of $\mathcal{F}$ is a total constructive function:
semi-decide in dovetail $x\in B_i$ for all $f_\uple{i,j}\in E$, and enumerate $p_j$ each time a test stops.

Given a total constructive function $f$, there is an algorithm $\A$ which on each $x\in X$ is extensional and describes $f(x)$, so $f=f_\A$.
\end{proof}

The proof even shows that the equivalence is constructive: the evaluation
  of any $f:X\to Y$ on any $x\in X$ can be achieved by an algorithm having
  access to any description of $f\in\C(X,Y)$, and any algorithm evaluating
  $f$ can be converted into an algorithm describing $f\in C(X,Y)$. More precisely:

\begin{proposition}\label{curry}Let $X,X'$ be computable metric spaces and $Y$ be an enumerative lattice:

\defin{Evaluation:} The function $Eval:\C(X,Y)\times X \to Y$ is constructive,

\defin{Curryfication:} If a function $f:X'\times X\to Y$ is constructive
then the function from $X'$ to $C(X,Y)$ mapping $x'\in X'$ to $f(x',.)$ is constructive.
\end{proposition}

Lemma \ref{lemma_extensional} and proposition
\ref{proposition_constructive_functions} implie:

\begin{corollary}\label{corollary_x_constructive}
The $x$-constructive elements of $Y$ are exactly the images of $x$ by \emph{total}
constructive functions from $X$ to $Y$. 
\end{corollary}

This is a particular property of the enumerative lattice structure: a partial constructive
function from some represented space to another cannot in general be
extended to a total constructive one.

\subsection{The Open Subsets of a computable metric space}\label{opensets}
Following \cite{BraWei99}, \cite{BraPre03}, we define constructivity notions on the open subsets of a computable metric space. The topology $\tau$ induced by the metric has the numbered set $\B$ of
ideal balls as a countable basis: any open set can then be described as a
countable union of ideal balls. Actually $(\tau,\subseteq,\B)$ is an
enumerative space (cf section \ref{section_es}), the supremum operator being union. The canonical representation on enumerative lattices (definition \ref{representation_es}) induces constructivity notions on $\tau$, a constructive open set being called a \defin{recursively enumerable (r.e) open set}.



On the integers, it may be unnatural to show that some subset is
recursively enumerable, and the equivalent notion of semi-decidable set is
often used. This notion can be extended to subsets of a computable metric space, and it happens
to be very useful in the applications. We recall from section
\ref{section_es} that $\{\bot,\top\}$ is an enumerative lattice, which induces canonically
the enumerative lattice $\C(X,\{\bot,\top\})$.

\begin{definition}
A subset $A$ of $X$ is said to be \defin{semi-decidable} if its
indicator function $1_A:X\to\{\bot,\top\}$ (mapping $x\in A$ to $\top$ and
$x\notin A$ to $\bot$) is constructive.
\end{definition}

In other words, $A$ is semi-decidable if there is a recursive function $\FI$ such that for all $x\in X$ and all description $\seq{s}$ of $x$, $\FI\orcl{\seq{s}}$ stops if and only if $x\in A$. It is a well-known result (see \cite{BraPre03}) that the two notions are effectively equivalent:

\begin{proposition}\label{proposition_semi-decidable}
A subset of $X$ is semi-decidable if and only if it is a r.e open set. Moreover, the enumerative lattices $(\tau,\subseteq,\B)$ and $\C(X,\{\bot,\top\})$ are constructively isomorphic.
\end{proposition}

The isomorphism is the function $U\mapsto 1_U$ and its inverse $f\mapsto
f^{-1}(\top)$. In other words, $f^{-1}(\top)$ is $f$-r.e uniformly in $f$
and $1_U$ is $U$-lower semi-computable uniformly in $U$. It implies in particular that:
\begin{corollary}
The intersection $(U,V)\mapsto U\cap V$ and union $(U,V)\mapsto U\cup V$
are constructive functions from $\tau\times \tau$ to $\tau$.
\end{corollary}

For computable functions between computable metric spaces, we have the following useful
characterization:

\begin{proposition}\label{functions}
Let $(X,d_X,S_X)$ and $(Y,d_Y,S_Y)$ be computable metric spaces. A function $f:X \vers Y$ is
computable on $D\subseteq X$ if and only if the preimages of ideal balls
are uniformly r.e open (in $D$) sets. That is, for all $i$,
$f^{-1}(B_i)=U_i \cap D$ where $U_i$ is a r.e open set uniformly in $i$.
\end{proposition}

We will use the following notion:
\begin{definition}
A $\Pi_2^0$-set is a set of the form $\bigcap_n U_n$ where $(U_n)_n$ is a
sequence of uniformly r.e open sets.
\end{definition}

\section{Computing with probability measures}\label{measures} 

\subsection{Measures as points of the computable metric space $\M(X)$}
\label{measures_section}

Here, following \cite{Gac05}, we define computable measures in the following way: first the space $\M(X)$ is endowed with a computable metric space structure compatible with the weak topology and then computable measures are defined as the constructive points.

Given a metric space $(X,d)$, the set $\M(X)$ of Borel probability measures over
$X$ can be endowed with the \emph{weak topology}, which is the finest
topology for which $\mu_n\vers\mu$ if and only if $\int f d\mu_n \vers \int
f d\mu$ for all continuous bounded function $f:X\to\R$. This topology is
metrizable and when $X$ is
separable and complete, $\M(X)$ is also separable and complete (see
\cite{Bil68}). Moreover, a computable metric structure on $X$ induces in a
canonical way a computable metric structure on $\M(X)$.

Let $\D\subset \M(X)$ be the set of those probability measures that
are concentrated in finitely many points of $\S$ and assign rational
values to them. It can be shown that this is a dense subset
(\cite{Bil68}). The numberings $\nu_{\S}$ of ideal points of $X$ and
$\nu_{\Q}$ of the rationals numbers induce a numbering $\nu_{\D}$ of ideal
measures: $\mu_{\uple{\uple{n_1,\ldots,n_k},\uple{m_1,\ldots,m_k}}}$ is the
measure concentrated over the finite set $\{s_{n_1},\ldots,s_{n_k}\}$ where
$q_{m_i}$ is the weight of $s_{n_i}$.

\subsubsection{The Prokhorov metric}

Let us consider the particular metric on $\M(X)$:

\begin{definition} The \defin{Prokhorov metric} $\rho$ on $\M(X)$ is defined by:
\begin{equation}\label{prokhorov}
\rho(\mu,\nu):=\inf \{\epsilon \in \R^+ : \mu(A)\leq\nu(A^{\epsilon})+\epsilon \mbox{ for every Borel set }A\}.
\end{equation}
where $A^{\epsilon}=\{x:d(x,A)< \epsilon\}$.
\end{definition}

It is known that it is indeed a metric, which induces the weak topology on
$\M(X)$ (see \cite{Bil68}). Moreover, we have that:

\begin{proposition}\label{prokhorov_ems}
 $(\M(X),\D,\rho)$ is a computable metric space.
\end{proposition}

\begin{proof}We have to show that the real numbers $\rho(\mu_i,\mu_j)$ are
  all computable, uniformly in $\uple{i,j}$. First observe that if $U$ is a
  r.e open subset of $X$, $\mu_i(U)$ is lower semi-computable uniformly in
  $i$ and $U$. Indeed, if $(s_{n_1},q_{m_1}),\ldots,(s_{n_k},q_{m_k})$ are
  the mass points of $\mu_i$ together with their weights (recoverable
  from $i$) then $\mu_i(U)=\sum_{s_{n_j}\in U} q_{m_j}$. As the $s_{n_j}$
  which belong to $U$ can be enumerated from any description of $U$, this sum is
  lower-semi-computable. In particular, $\mu_i(B_{i_1}\cup\ldots\cup
  B_{i_k})$ is lower semi-computable and
  $\mu_i(\overline{B}_{i_1}\cup\ldots\cup \overline{B}_{i_k})$ is upper
  semi-computable, both of them uniformly in $\uple{i,i_1,\ldots,i_k}$

Now we prove that $\rho(\mu_i,\mu_j)$ is computable uniformly in $\uple{i,j}$. 

Observe that if $\mu_i$ is an ideal measure concentrated over $S_i$, then (\ref{prokhorov}) becomes $\rho(\mu_i,\mu_j)=\inf \{\epsilon\in\Q:\forall A \subset S_i\mbox{, } \mu_i(A)<\mu_j(A^{\epsilon})+\epsilon\}$. Since $\mu_j$ is also an ideal measure and $A^{\epsilon}$ is a finite union of open ideal balls, the number $\mu_j(A^{\epsilon})$ is lower semi-computable (uniformly) and then $\rho(\mu_i,\mu_j)$ is upper semi-computable, uniformly in $\uple{i,j}$. To see that $\rho(\mu_i,\mu_j)$ is lower-semicomputable, uniformly in $\uple{i,j}$, observe that $\rho(\mu_i,\mu_j)=\sup \{\epsilon\in\Q:\exists A \subset S_i\mbox{, } \mu_i(A)>\mu_j(A^{\overline{\epsilon}})+\epsilon\}$, where $A^{\overline{\epsilon}}=\{x:d(x,A)\leq\epsilon\}$ (a finite union of closed ideal balls when $A\subset S_i$) and use the upper semi-computability of $\mu_j(A^{\overline{\epsilon}})$.
\end{proof}

\begin{definition}\label{computableMu}
A measure $\mu$ is \defin{computable} if it is a constructive point of $(\M(X),\D,\rho)$.
\end{definition}

The effectivization of the space of Borel probability measures $\M(X)$ is
of theoretical interest, and opens the question: what kind of information can be
(algorithmically) recovered from a description of a measure as a point of
the computable metric space $\M(X)$ ? The two most current uses of a measure are to give weights
to measurable sets and means to measurable functions. Can these quantities
be computed ?

\subsubsection{The Wasserstein metric}

In the particular case when the metric space $X$ is bounded, an alternative
metric can be defined on $\M(X)$. When $f$ is a real-valued function, $\mu f$
denotes $\int f d\mu$.

\begin{definition}
The \textit{Wasserstein metric} on $\M(X)$ is defined by:
\begin{equation}\label{Wass}
W(\mu,\nu)=\sup_{f\in 1-Lip(X)}(|\mu f-\nu f|)
\end{equation}
where $1-Lip(X)$ is the space of 1-Lipschitz functions from $X$ to $\R$.
\end{definition}

We recall (see \cite{AmbGigSav05}) that $W$ has the following properties:
\begin{proposition} $\quad$
\begin{enumerate}
 \item $W$ is a distance and if $X$ is separable and complete then $\M(X)$ with this distance is a separable and complete metric space.
 \item The topology induced by $W$ is the weak topology and thus $W$ is equivalent to the Prokhorov metric.
\end{enumerate}
\end{proposition}

Moreover, if $(X,\S,d)$ is a computable metric space (and $X$ bounded), then:

\begin{proposition}
$(\M(X),\D,W)$ is a computable metric space.
\end{proposition}

\begin{proof}
We have to show that the distance $W(\mu_i,\mu_j)$ between ideal measures is uniformly computable. From $\uple{i,j}$ we can compute the set $S_{i,j}=supp(\mu_i)\cup supp(\mu_j)$. Let $s_0\in S_{i,j}$, then we can suppose that the supremum in (\ref{Wass}) is taken over $1-Lip^0_{s_0}(X):=\{f\in 1-Lip(X): 1-Lip^0_{s_0}(X)f(s)=0\}$. Given some precision $\epsilon$ we construct a finite set $\mathcal{N}_{\epsilon}\subset 1-Lip^0_{s_0}(X)$ made of uniformly computable functions such that for each $f\in 1-Lip^0_{s_0}(X)$ there is some $l\in \mathcal{N}_{\epsilon}$ satisfying $\sup \{|f(x)-l(x)|:x\in S_{i,j}\}<\epsilon$: compute an integer $m$ such that $S_{i,j}\subset B(s,m)$; then $|f|<m$ for every $f\in 1-Lip^0_s(X)$. Let $n$  be such that $m/n<2\epsilon$. For each $s\in S_{i,j}$ and $a\in \{\frac{lm}{n}\}_{l=-m}^m$ let us consider the functions defined by $\phi^+_{s,l}(x):=a+d(s,x)$ and $\phi^-_{s,l}(x):=a-d(s,x)$. Then it is not difficult to see that $\mathcal{N}_{\epsilon}$ defined as the set of all possible combinations of $\max$ and $\min$ made with the $\phi^{+-}_{s,l}(x)$ satisfy the required condition. 

Therefore, since $\sup(|f-g|)<\epsilon$ implies $|\mu (f-g)|<\epsilon$ we have that: 
$$
W(\mu_i,\mu_j)\in [\sup_{g\in \mathcal{N}_{\epsilon}}(|\mu_i g - \mu_j g|),\sup_{g\in \mathcal{N}_{\epsilon}}(|\mu_i g - \mu_j g|)+2\epsilon]
$$

where the $\mu_i g$ are computable, uniformly in $i$. The result follows.
\end{proof}

When $X$ is bounded, the effectivisation using the Prokhorov or the
Wasserstein metrics turn out to be equivalent.

\begin{theorem}The Prokhorov and the Wasserstein metrics are computably
  equivalent. That is, the identity function $id:(\M(X),\D,\rho)\vers (\M(X),\D,W)$
  is a computable isomorphism, as well as its inverse.
\end{theorem}

\begin{proof}
Let $M$ be an integer such that $\sup_{x,y\in X}d(x,y)<M$. Suppose $\rho(\mu,\nu)<\epsilon/(M+1)$. Then, by the coupling theorem \cite{Bil68}, for every $f\in 1-Lip(X)$ it holds $|\mu f-\nu f|\leq \epsilon$, then $W(\mu,\nu)<\epsilon$. Conversely, suppose $W(\mu,\nu)<\epsilon^2<1$. Let $A$ be a Borel set and define $g_{\epsilon}^A:=|1-d(x,A)/\epsilon|^+$. Then $\epsilon g_{\epsilon}^A \in 1-Lip(X)$. $W(\mu,\nu)<\epsilon^2$ implies $\mu \epsilon g_{\epsilon}^A < \nu \epsilon g_{\epsilon}^A + \epsilon^2 $ and since  $\mu(A)\leq\mu g_{\epsilon}^A$ and $\nu g_{\epsilon}^A\leq \nu(A^{\epsilon})$, we conclude $\mu(A)\leq \nu(A^{\epsilon})+\epsilon$ and then $\rho(\mu,\nu)<\epsilon$. Therefore, given a fast sequence of ideal measures converging to $\mu$ in the Prokhorov metric, we can construct a fast sequence of ideal measures converging to $\mu$ in the $W$ metric and vice-versa.
\end{proof}

This equivalence offers an alternative method to prove computability of measures. It is used for example in \cite{GalHoyRoj07a} to show the computability  of the physical measures for some classes of dynamical systems.

\subsection{Measures as valuations}

We now investigate the first problem: can the measure of sets be computed
from the Cauchy description of a measure? Actually, the answer is positive
for a very small part of the Borel sigma-field. It is a well-known fact that a Borel (probability) measure $\mu$ is characterized by the measure of open sets, which generate the Borel sigma-field. That is, by the
valuation $v_\mu:\tau\to[0,1]$ which maps an open set to its
$\mu$-measure. The question is then so study this characterization from a
computability viewpoint.



The first result is that the measure of open sets can be lower
semi-computed, using the Cauchy description of the measure.

\begin{proposition}\label{val_operator}
The valuation operator $v:\M(X)\times \tau \to [0,1]$ mapping $(\mu,U)$ to $\mu(U)$ is lower semi-computable. 
\end{proposition}

\begin{proof}
as $v_\mu=v(\mu,.)$ is Scott-continuous (see remark \ref{scott}), it suffices to show that it is
uniformly lower semi-computable on finite unions of balls.

We first restrict to ideal measures $\mu_i$: we have already seen (proof of
proposition \ref{prokhorov_ems}) that all
$\mu_i(B_{i_1}\cup\ldots\cup B_{i_k})$ are lower semi-computable real
numbers, uniformly in $\uple{i,i_1,\ldots,i_k}$.

Now let $(\mu_{k_n})_{n\in\N}$ a description of a measure $\mu$, that is a fast
sequence converging to $\mu$ for the Prokhorov distance: then
$\rho(\mu_{k_n},\mu)\leq \epsilon_n$ where $\epsilon_n=2^{-n+1}$. For
$n\geq 1$, and $U=B(s_{i_1},q_{j_1})\cup\ldots\cup B(s_{i_k},q_{j_k})$
define:
$$
U_n=\bigcup_{m\leq k} B(s_{i_m},q_{j_m}-\epsilon_n)
$$
note that
$U_{n-1}^{\epsilon_n}\subseteq U_n$ and $U_n^{\epsilon_n}\subseteq U$. We show that $\mu(U)=\sup_n (\mu_{j_n}(U_n)-\epsilon_n)$:

\puce $\mu_{j_n}(U_n)\leq \mu(U)+\epsilon_n$ for all $n$, so
$\mu(U)\geq \sup_n (\mu_{j_n}(U_n)-\epsilon_n)$.

\puce $\mu(U_{n-1})\leq \mu_{j_n}(U_n)+\epsilon_n$ for all $n$. As
 $U_{n-1}$ increases towards $U$ as $n\to\infty$, $\mu(U)=\sup_n (\mu(U_{n-1})-2\epsilon_n)\leq \sup_n (\mu_{j_n}(U_n)-\epsilon_n)$.

As the quantity $\mu_{j_n}(U_n)-\epsilon_n$ is lower semi-computable
uniformly in $n$, we are done (observe that everything is uniform in the
finite description of $U$).
\end{proof}

The second result is stronger: the lower semi-computability of the measure of
the r.e open sets even characterizes the computability of the measure.

\begin{theorem}\label{valuation_equivalence} Given a measure $\mu\in\M(X)$, the following are equivalent:
\begin{enumerate}
\item $\mu$ is computable,
\item $v_{\mu}:\tau\to[0,1]$ is lower-semi-computable,
\item $\mu(B_{i_1}\cup\ldots\cup B_{i_k})$ is lower-semi-computable uniformly in $\uple{i_1,\ldots,i_k}$.
\end{enumerate}
\end{theorem}

\begin{proof}

$[1\Rightarrow 2]$ Direct from proposition \ref{val_operator}. $[2\Rightarrow 3]$
  Trivial. $[3\Rightarrow 1]$ We show that $\rho(\mu_n,\mu)$ is upper semi-computable
  uniformly in $n$, and then use proposition \ref{proposition_computable_point}. Since
  $\rho(\mu_n,\mu)<\epsilon$ iff $\mu_n(A)<\mu(A^{\epsilon})+\epsilon$ for
  all $A\subset S_n$ where $S_n$ is the finite support of $\mu_n$, and
  $\mu(A^{\epsilon})$ is lower semi-computable ($A^{\epsilon}$ is a finite
  union of open ideal balls) $\rho(\mu_n,\mu)<\epsilon$ is semi-decidable,
  uniformly in $n$ and $\epsilon$. This allows to construct a fast
  sequence of ideal measures converging to $\mu$.
\end{proof}

It means that a representation which would be ``tailor-made'' to make the
valuation constructive, describing a measure $\mu$ by the set of integers
$\uple{i_1,\ldots,i_k,j}$ satisfying $\mu(B_{i_1}\cup\ldots\cup
B_{i_k})>q_j$, would be constructively equivalent to the Cauchy
representation. This is the approach taken in \cite{Wei99} for the special
case $X=[0,1]$ and in \cite{Sch07} on an arbitrary sequential topological
space. In both case, the topology on $\M(X)$ induced by this representation
is proved to be equivalent to the weak topology. A domain theoretical
approach was also developed in \cite{Eda96} on a compact
space, the Scott topology being proved to induce the weak topology.

\subsubsection{The examples of the Cantor space and the unit interval.}

On the Cantor space $\Sigma^\N$ (where $\Sigma$ is a finite alphabet) with
its natural computable metric space structure, the ideal balls are the cylinders. As a finite
union of cylinders can always be expressed as a disjoint (and finite) union of
cylinders, and the complement of a cylinder is a finite union of cylinders,
we have:

\begin{corollary}\label{measure_cantor}
A measure $\mu \in \M(\Sigma^{\N})$ is computable iff the measures of
the cylinders are uniformly computable.
\end{corollary}

On the unit real interval, ideals balls are open rational intervals. Again, a
finite union of such intervals can always be expressed as a disjoint
(and finite) union of open rational intervals. Then:

\begin{corollary}A measure $\mu \in \M([0,1])$ is computable iff the
  measures of the rational open intervals are uniformly lower-semi-computable.
\end{corollary}

If $\mu$ has no atoms, a rational open interval is the complement of at
most two disjoint open rational intervals, up to a null set. In this case,
$\mu$ is then computable iff the measures of the
rational intervals are uniformly \emph{computable}.

\subsection{Measures as integrals}

We now answer the second question: is the integral of functions
computable from the description of a measure ?

The computable metric space structure of $X$ and the enumerative lattice structure of $\Rplus$ induce
in a canonical way the enumerative space $\C(X,\Rplus)$ (see
section \ref{section_CXY}), which is actually the set of lower
semi-continuous functions from $X$ to $\Rplus$. We have:

\begin{proposition}\label{int_operator}
The integral operator $\int:\M(X)\times\C(X,\Rplus)\to\Rplus$ is lower semi-computable.
\end{proposition}

\begin{proof}
the integral of a finite supremum of step functions can be expressed by
induction on the number of functions: first, $\int
f_\uple{i,j}d\mu=q_j\mu(B_i)$ and
\begin{eqnarray*}
\int\sup\{f_\uple{i_1,j_1},\ldots,f_\uple{i_k,j_k}\}d\mu & = & q_{j_m}\mu(B_{i_1}\cup\ldots\cup
B_{i_k})+{} \\
& & \int\sup\{f_\uple{i_1,j'_1},\ldots,f_\uple{i_k,j'_k}\}d\mu
\end{eqnarray*}
where $q_{j_m}$ is minimal among $\{q_{j_1},\ldots,q_{j_k}\}$ and $q_{j'_1}=q_{j_1}-q_{j_m},
q_{j'_2}=q_{j_2}-q_{j_m}, etc$. Note that $f_\uple{i_m,j'_m}$ being the zero
function can be removed.

Now, $m$ can be computed and by proposition \ref{val_operator} the measure of finite unions of ideal balls can
be uniformly $\mu$-lower semi-computed, so the integral above
can be uniformly $\mu$-lower semi-computed. For any fixed measure $\mu$, the
integral operator $\int d\mu:\C(X,\Rplus)\to\Rplus$ is Scott-continuous,
so it is lower semi-computable.
\end{proof}

Again, the lower semi-computability of the integral of lower
semi-computable functions characterizes the computability of the measure:

\begin{corollary}Given a measure $\mu\in\M(X)$, the following are equivalent:
\begin{enumerate}
\item $\mu$ is computable,
\item $\int d\mu:\C(X,\Rplus)\to\Rplus$ is lower semi-computable,
\item $\int \sup\{f_{i_1},\ldots,f_{i_k}\}d\mu$ is lower-semi-computable uniformly in $\uple{i_1,\ldots,i_k}$.
\end{enumerate}
\end{corollary}

\begin{proof}

$[2\Leftrightarrow 3]$ holds by Scott-continuity of the operator,

$[1\Rightarrow 2]$ is a direct consequence of proposition
  \ref{int_operator},

$[2\Rightarrow 1]$ is a direct consequence of theorem \ref{valuation_equivalence},
  composing the integral operator with the function from
  $\tau$ to $\C(X,\Rplus)$ mapping an open set to its indicator function
  (which is computable, see proposition \ref{proposition_semi-decidable}).
\end{proof}

It means that a representation of measures which would be ``tailor-made''
to make the integration constructive, describing a measure by the set of
integers $\uple{i_1,\ldots,i_k,j}$ satisfying
$\int\sup\{f_{i_1},\ldots,f_{i_k}\}d\mu> q_j$, would be constructively
equivalent to the Cauchy representation.

A corollary of proposition \ref{int_operator} will be used in the last
section: let $(f_i)_i$ be a sequence of uniformly computable functions, i.e. such
that the function $(i,x)\mapsto f_i(x)$ is computable. If moreover $f_i$
has a bound $M_i$ computable uniformly in $i$, then the function
$(\mu,i)\to \int f_id\mu$ is computable. Indeed, $f_i+M_i$
(resp. $M_i-f_i$) is uniformly lower (resp. upper) semi-computable, so
$\int f_id\mu= \int (f_i+M_i)d\mu -M_i=M_i-\int (M_i-f_i)d\mu$ and
proposition \ref{int_operator} allow to conclude.

\section{Computable Probability Spaces}\label{cps}

The representation induced by the binary numeral system of real numbers is generally presented as not adequate for computability purposes since simple functions as $x \mapsto 3x$ are not computable with respect to it. This lies in the fact that the real interval and the space of sequences are not homeomorphic. 

On the other hand, if we are insterested in probabilistic issues, the binary representation is actually suitable, and may even be preferred: almost every real has a unique binary expansion.

More generally, computability notions from computable analysis are effective versions of topological ones (semi-decidable sets are open, computable functions are continuous, etc). What about effective versions of measure-theoretical/probabilistic notions? 

In this section we study a computable version of probability spaces, that is, metric spaces equipped with a fixed computable Borel probability measure. This will give us a framework allowing to talk about \emph{almost everywhere} computability or decidability notions. Let us then introduce:

\begin{definition}
A \defin{computable probability space} is a pair $(\X,\mu)$  where
$\X$ is a computable metric space and $\mu$ a computable Borel probability measure on $X$.
\end{definition}


\begin{definition}
A \defin{morphism of computable probability spaces} $F:(\X,\mu)\to(\mathcal{Y},\nu)$, is a
computable measure-preserving function $F:D_F\subseteq X\to Y$ where $D_F$ is
a (full-measure) $\Pi^0_2$-set.

An \defin{isomorphism} $(F,G):(\X,\mu)\rightleftarrows(\mathcal{Y},\nu)$ is a pair $(F,G)$ of morphisms such that $G\circ F=id$ on $F^{-1}(D_G)$ and $F\circ G=id$ on $G^{-1}(D_F)$.
\end{definition}

We recall that $F$ is measure-preserving if $\nu(A)=\mu(F^{-1}(A))$ for all
Borel set $A$.






\subsection{Generalized binary representations}

The Cantor space $2^\omega$ ($2$ denotes $\{0,1\}$) is a privileged place
for computability. This can be understood by the fact that it is the
countable product (with the product topology) of a finite space (with the
discrete topology). A consequence of this is that membership of a basic
open set (cylinder) boils down to a pattern-matching and is then
\emph{decidable}. As decidable sets must be clopen, this property cannot
hold in connected spaces. As a result, a computable metric space is not in
general constructively homeomorphic to the Cantor space.

Nevertheless, the real unit interval $[0,1]$ is not so far away from the
Cantor space. The binary numeral system provides a correspondence between
real numbers and binary sequences, which is certainly not homeomorphic,
unless we remove the small set of dyadic numbers. In particular, the
remaining set is totally disconnected, and the dyadic intervals form a
basis of clopen sets.

Actually, this correspondence makes the computable probability space
$[0,1]$ with the Lebesgue measure isomorphic to the Cantor space with the
uniform measure. This fact has been implicitly used, for instance, to
extend algorithmic randomness from the Cantor space with the uniform measure
to the unit interval with the Lebesgue measure.

We generalize this to any computable probability space, over which we define the notion of \emph{binary
representation}. We show that every computable probability space has a binary representation. This implies, in particular, that every computable probability space is isomorphic to the Cantor space with
a computable measure. To carry out this generalization, let us briefly scrutinize
the binary numeral system on the unit interval:



$\delta:2^\omega\to [0,1]$ is a total surjective morphism. Every non-dyadic
real has a unique expansion, and the inverse of $\delta$, defined on the
set $D$ of non-dyadic numbers, is computable. Moreover, $D$ is large both
in a topological and measure-theoretical sense: it is a residual (a
countable intersection of dense open sets) and has measure
one. $(\delta,\delta^{-1})$ is then an isomorphism.

In our generalization, we do not require every binary sequence to be the
expansion of a point, which would force $X$ to be compact.

\begin{definition}\label{binary} A \defin{binary representation} of
a computable probability space $(\X,\mu)$ is a pair $(\delta,\mu_\delta)$
where $\mu_\delta$ is a computable probability measure on $2^\omega$ and
$\delta:(2^\omega,\mu_\delta)\to(\X,\mu)$ is a surjective morphism such
that, calling $\delta^{-1}(x)$ the set of \defin{expansions} of $x\in X$:

\puce there is a dense
full-measure $\Pi^0_2$-set $D$ of points having a unique expansion,

\puce $\delta^{-1}:D\to\delta^{-1}(D)$ is computable.
\end{definition}

Remark that when the support of the measure (the smallest closed set of
full measure) is the whole space $X$, like the Lebesgue
measure on the interval, a full-measure $\Pi^0_2$-set is always a residual,
but in general it is only dense on the support of the measure: that is the
reason why we explicitly require $D$ to be dense. Also remark that a
binary representation $\delta$ always induces an isomorphism
$(\delta,\delta^{-1})$ between the Cantor space and the computable probability space.



The sequel of this section is devoted to the proof of the following result:

\begin{theorem}\label{theorem_binary}
Every computable probability space $(\X,\mu)$ has a binary representation.
\end{theorem}

The space, restricted to the domain $D$ of the isomorphism, is then
totally disconnected: the preimages of the cylinders form a basis of
clopen and even \emph{decidable} sets. In the whole space, they are not
decidable any more. Instead, they are \emph{almost decidable}.

\begin{definition}
A set $A$ is said to be \defin{almost decidable}
if there are two r.e open sets $U$ and $V$ such that:
$$
U\subset A, \quad V\subseteq \comp{A}, \quad U\cup V\mbox{ is dense and has
measure one}
$$
\end{definition}

\begin{definition}
A measurable set $A$ is said to be \defin{$\boldsymbol{\mu}$-continuous} or a \defin{$\boldsymbol{\mu}$-continuity set} if $\mu(\partial A)=0$ where $\partial A=\overline{A}\cap\overline{X\setminus A}$ is the boundary of $A$.
\end{definition}

Remark that, as for subsets of $\N$, a set is almost decidable if and only
if its complement is a.s. decidable. An almost decidable set is always a
continuity set. Let $B(s,r)$ be a $\mu$-continuous ball with computable
radius: in general it is not an almost decidable set (for instance, isolated points may
be at distance exactly $r$ from $s$). But if there is no ideal point is at distance $r$
from $s$, then $B(s,r)$ is almost decidable: take $U=B(s,r)$ and
$V=X\setminus \overline{B}(s,r)$.

We say that the elements of a sequence $(A_i)_{i\in\N}$ are uniformly
a.s. decidable if there are two sequences $(U_i)_{i\in\N}$ and
$(V_i)_{i\in\N}$ of uniformly r.e sets satisfying the conditions above.

\begin{lemma}
There is a sequence $(r_n)_{n\in\N}$ of uniformly computable reals such
that $(B(s_i,r_n))_\uple{i,n}$ is a basis of uniformly almost decidable
balls.
\end{lemma}

\begin{proof}
define $U_\uple{i,k}=\{r\in \R^+:
\mu(\overline{B}(s_i,r))<\mu(B(s_i,r))+1/k\}$: by computability of $\mu$,
this is a r.e open subset of $\R^+$, uniformly in $\uple{i,k}$. It is
furthermore dense in $\R^+$: the spheres $S_r=\overline{B}(s_i,r)\setminus
B(s_i,r)$ form a partition of the space when $r$ varies in $\R^+$ and $\mu$
is finite, so the set of $r$ for which $\mu(S_r)\geq 1/k$ is finite.

Define $V_\uple{i,j}=\R^+\setminus\{d(s_i,s_j)\}$: this is a dense r.e open set,
uniformly in $\uple{i,j}$.

Then by the computable Baire Category Theorem (see \cite{YasMorTsu99},
\cite{Bra01}), the dense $\Pi^0_2$-set $\bigcap_\uple{i,k} U_\uple{i,k}\cap
\bigcap_\uple{i,j} V_\uple{i,j}$ contains a sequence $(r_n)_{n\in\N}$ of uniformly computable real numbers
which is dense in $\R^+$. In other words, all $r_n$ are computable,
uniformly in $n$. By construction, for any $s_i$ and $r_n$, $B(s_i,r_n)$ is
almost decidable.

We recall that from an enumeration $(I_n)_{n\in\N}$ of all the rational
compact intervals of $\R^+$, $r_n$ is constructed computing a nested
shrinking sequence $(J^n_k)_{k\in\N}$ of rational compact intervals
starting from $J^n_0=I_n$, and such that $J^n_{k+1}\subseteq J^n_k\cap
U_k\cap V_k$. Then $\{r_n\}=\bigcap_k J^n_k$.
\end{proof}

We will denote $B(s_i,r_n)$ by $B^\mu_k$ where $k=\uple{i,n}$. Note that
different algorithmic descriptions of the same $\mu$ may yield different
sequences $(r_n)_{n\in\N}$, so $B^\mu_k$ is an abusive notation. It is
understood that some algorithmic description of $\mu$ has been chosen and
fixed. This can be done only because the measure $\mu$ is computable, which
is then a crucial hypothesis. We denote $X\setminus \overline{B}(s_i,r_n)$ by $C^\mu_k$ and define:

\begin{definition}
For $w\in 2^*$, the \defin{cell} $\Gamma(w)$ is defined by induction on $|w|$:

$$
\Gamma(\epsilon)=X, \quad \Gamma(w0)=\Gamma(w)\cap C^\mu_i \quad \mbox{ and
} \quad 
\Gamma(w1)=\Gamma(w)\cap B^\mu_i
$$
where $\epsilon$ is the empty word and $i=|w|$.
\end{definition}
This an almost decidable set, uniformly in $w$. \\

\begin{proof}(of theorem \ref{theorem_binary}). We construct an encoding function $b:D\to 2^\omega$, a decoding
function $\delta:D_\delta\to X$, and show that $\delta$ is a binary
representation, with $b=\delta^{-1}$.

\textbf{Encoding.}

Let $D=\bigcap_i B^\mu_i \cup C^\mu_i$: this is a dense
full-measure $\Pi^0_2$-set. Define the computable function $b:D\to 2^\omega$ by:
$$
b(x)_i=\left\{\begin{array}{rl}
1 & \mbox{ if } x\in B^\mu_i \\
0 & \mbox{ if } x\in C^\mu_i
\end{array}\right.
$$

Let $x\in D$: $\omega=b(x)$ is also characterized by $\{x\}=\bigcap_i
\Gamma(\omega_{0..i-1})$. Let $\mu_\delta$ be the image measure of $\mu$ by $b$:
$\mu_\delta=\mu\circ b^{-1}$. $b$ is then a morphism from $(X,\mu)$ to $(2^\omega,\mu_\delta)$.

\textbf{Decoding.}

Let $D_\delta$ be the set of binary sequences $\omega$ such that $\bigcap_i
\overline{\Gamma(\omega_{0..i-1})}$ is a singleton. We define the decoding
function $\delta:D_\delta\to X$ by:
$$
\delta(\omega)=x \mbox{ if } \bigcap_i \overline{\Gamma(\omega_{0..i-1})}=\{x\}
$$

$\omega$ is called an expansion of $x$. Remark that $x\in
B^\mu_i\Rightarrow \omega_i=1$ and $x\in C^\mu_i\Rightarrow \omega_i=0$, which
implies in particular that if $x\in D$, $x$ has a unique expansion, which
is $b(x)$. Hence, $b=\delta^{-1}:\delta^{-1}(D)\to D$ and $\mu_\delta(D_\delta)=\mu(D)=1$.

We now show that $\delta:D_\delta \to X$ is a surjective morphism. For seek
of clarity, the center and the radius of the ball $B^\mu_i$ will be denoted
$s_i$ and $r_i$ respectively. Let us call $i$ an $n$-\emph{witness} for
$\omega$ if $r_i<2^{-(n+1)}, \omega_i=1$ and $\Gamma(\omega_{0..i})\neq
\emptyset$. 

\puce $D_\delta$ is a $\Pi^0_2$-set: we show that $D_\delta=\bigcap_n
\{\omega\in 2^\omega:\omega \mbox{ has a $n$-witness}\}$.

Let $\omega\in D_\delta$ and $x=\delta(\omega)$. For
each $n$, $x\in B(s_i,r_i)$ for some $i$ with $r_i<2^{-(n+1)}$. Since
$x\in\overline{\Gamma(\omega_{0..i})}$, we have that $\Gamma(\omega_{0..i})\neq
\emptyset$ and $\omega_i=1$ (otherwise $\overline{\Gamma(\omega_{0..i})}$ is disjoint of $B^\mu_i$). In other words,
$i$ is an $n$-witness for $\omega$.

Conversely, if $\omega$ has a $n$-witness $i_n$ for all $n$,
since $\overline{\Gamma(\omega_{0..i_n})}\subseteq \overline{B^\mu_{i_n}}$ whose radius
tends to zero, the nested sequence
$(\overline{\Gamma(\omega_{0..i_n})})_n$ of closed cells has, by
completeness of the space, a non-empty intersection, which is a singleton.

\puce $\delta:D_\delta\to X$ is computable. For each $n$, find some
$n$-witness $i_n$ of $\omega$: the sequence $(s_{i_n})_n$ is a fast
sequence converging to $\delta(\omega)$.

\puce $\delta$ is surjective: we show that each point $x\in X$
has at least one expansion. To do this, we construct by induction a
sequence $\omega=\omega_0\omega_1\ldots$ such that for all $i$,
$x\in\overline{\Gamma(\omega_0\ldots \omega_i)}$. Let $i\geq 0$ and suppose
that $\omega_0\ldots \omega_{i-1}$ (empty when $i=0$) has been
constructed. As $B^\mu_i\cup C^\mu_i$ is open dense and
$\Gamma(\omega_{0..i-1})$ is open,
$\overline{\Gamma(\omega_{0..i-1})}=\overline{\Gamma(\omega_{0..i-1})\cap
(B^\mu_i\cup C^\mu_i)}$ which equals $\overline{\Gamma(\omega_{0..i-1}0)}\cup
\overline{\Gamma(\omega_{0..i-1}1)}$. Hence, one choice for
$\omega_i\in\{0,1\}$ gives $x\in
\overline{\Gamma(\omega_{0..i})}$.

By construction, $x\in \bigcap_i\overline{\Gamma(\omega_{0..i-1})}$. As
$(B^\mu_i)_i$ is a basis and $\omega_i=1$ whenever $x\in B^\mu_i$ ,
$\omega$ is an expansion of $x$.
\end{proof}

\subsection{Another characterization of the computability of measures}
The existence of a basis of almost decidable sets also leads to another
characterization of the computability of measures, which is reminiscent of
what happens on the Cantor space (see corollary \ref{measure_cantor}). Let us
say that two bases $(U_i)_i$ and $(V_i)_i$ of the topology $\tau$ are constructively
equivalent if both $id_\tau:(\tau,\subseteq,\U)\to(\tau,\subseteq,\V)$ and
its inverse are constructive functions between enumerative lattices.

\begin{corollary}
A measure $\mu\in\M(X)$ is computable if and only if there is a basis
$\mathcal{U}=(U_i)_{i\in\N}$ of uniformly almost decidable open sets which is
constructively equivalent to $\B$ and such that all
$\mu(U_{i_1}\cup\ldots\cup U_{i_k})$ are computable uniformly in
$\uple{i_1,\ldots,i_k}$.
\end{corollary}

\begin{proof}
if $\mu$ is computable, the a.s. decidable balls $U_\uple{i,n}=B(s_i,r_n)$
are basis which is constructively equivalent to $\B$: indeed,
$B(s_i,r_n)=\bigcup_{q_j<r_n} B(s_i,q_j)$ and $B(s_i,q_j)=\bigcup_{r_n<q_j}
B(s_i,r_n)$, and $r_n$ is computable uniformly in $n$.

For the converse, the valuation function $f_\mu$ is lower
semi-computable. Indeed, the r.e open sets are uniformly r.e relatively to
the basis $\mathcal{U}$, so their measures can be lower-semi-computed,
computing the measures of finite unions of elements of $\mathcal{U}$. Hence
$\mu$ is computable by theorem \ref{valuation_equivalence}.
\end{proof}


\section{Algorithmic randomness}\label{randomness}

On the Cantor space with a computable measure $\mu$, Martin-L\"of
originally defined the notion of an individual random sequence as a
sequence passing all $\mu$-randomness tests. A $\mu$-randomness test
\emph{\`a la Martin-L\"of} is a sequence of uniformly r.e open sets
$(U_n)_n$ satisfying $\mu(U_n)\leq 2^{-n}$. The set $\bigcap_n U_n$ has
null measure, in an effective way: it is then called an effective null set.

Equivalently, a $\mu$-randomness test can be defined as a positive lower
semi-computable function $t:2^\omega\to\R$ satisfying $\int td\mu\leq 1$
(see \cite{VovVyu93} for instance). The associated effective null set is
$\{x:t(x)=+\infty\}=\bigcap_n \{x:t(x)>2^n\}$. Actually, every effective
null set can be put in this form for some $t$. A point is then called $\mu$-random if it lies in no effective null set.






Following G\'acs, we will use the second presentation of randomness tests which is
more suitable to express uniformity.

\subsection{Randomness w.r.t any probability measure}

\begin{definition}
Given a measure $\mu\in\M(X)$, a \defin{$\boldsymbol{\mu}$-randomness test} is a
$\mu$-constructive element $t$ of $\C(X,\Rplus)$, such that $\int
td\mu\leq 1$. Any subset of $\{x\in X: t(x)=+\infty\}$ is called a
\defin{$\boldsymbol{\mu}$-effective null set}.

A \defin{uniform randomness test} is a constructive function
$T$ from $\M(X)$ to $\C(X,\Rplus)$ such that for all $\mu\in\M(x)$, $\int
T^\mu d\mu\leq 1$ where $T^\mu$ denotes $T(\mu)$.
\end{definition}

Note that $T$ can be also seen as a lower-semi-computable function
from $\M(X)\times X$ to $\Rplus$ (see section \ref{section_CXY}).

A presentation \emph{\`a la Martin-L\"of} can be directly obtained using
the constructive functions $F: \C(X,\Rplus) \to \tau^\N$ and $G: \tau^\N \to \C(X,\Rplus)$ defined by: 
$F(t)_n := t^{-1}(2^n,+\infty)$ and $G((U_n)_n)(x):= \sup\{n:x\in \bigcap_{i\leq n} U_i\}$. They satisfy $F\circ G=id:\tau^\N\to\tau^\N$ and preserve the corresponding effective
null sets.

A uniform randomness test $T$ induces a $\mu$-randomness test
$T^\mu$ for all $\mu$. We show two important results which hold on any computable metric space:

\puce the two notions are actually equivalent (theorem \ref{theorem_uniform}), 

\puce there is a \emph{universal} uniform randomness test (theorem \ref{theorem_universal_test}).

The second result was already obtained by G\'acs, but only on spaces which have
recognizable Boolean inclusions, which is an additional computability property on the
basis of ideal balls.

By proposition \ref{proposition_constructive_functions}, constructive functions from $\M(X)$ to $\C(X,\Rplus)$ can be identified to constructive elements of the enumerative lattice $\C(\M(X),\C(X,\Rplus))$. Let $(H_i)_{i\in\N}$ be an enumeration of all its constructive elements (proposition \ref{enum_es}): $H_i=\sup_k
f_{\FI(i,k)}$ where $\FI:\N^2\to\N$ is some recursive function and the $f_n$ are step functions.

\begin{lemma}There is a constructive function $T:\N\times
  \M(X)\to\C(X,\Rplus)$ satisfying:

\puce for all $i$, $T_i=T(i,.)$ is a uniform randomness test,

\puce if $\int H_i(\mu)d\mu<1$ for some $\mu$, then $T_i(\mu)=H_i(\mu)$.
\end{lemma}

\begin{proof}
To enumerate only tests, we would like to be able to semi-decide $\int \sup_{k<n}f_{\FI(i,k)}(\mu)d\mu<1$. But $\sup_{k<n}f_{\FI(i,k)}(\mu)$ is only lower semi-computable (from $\mu$). To overcome this problem, we use another class of basic function.

Let $\Y$ be a computable metric space: for an ideal point $s$ of $Y$ and positive rationals $q,r,\epsilon$, define the \emph{hat} function:
$$
h_{q,s,r,\epsilon}(y):=q.[1-[d(y,s)-r]^+/\epsilon]^+
$$ where $[a]^+=\max\{0,a\}$. This is a continuous function whose value is
$q$ in $B(s,r)$, $0$ outside $B(s,r+\epsilon)$. The numberings of $\S$ and
$\Q_{>0}$ induce a numbering $(h_n)_{n\in\N}$ of all the hat functions. They can be taken as an alternative to step functions in the enumerative lattice $\C(Y,\Rplus)$: they yield the same computable structure. Indeed, step functions can be constructively expressed
as suprema of such functions: $f_\uple{i,j}=\sup \{h_{q_j,s,r-\epsilon,\epsilon}:
0<\epsilon<r\}$ where $B_i=B(s,r)$, and conversely.

We apply this to $Y=\M(X)\times X$ endowed with the canonical computable metric structure. By Curryfication it provides functions $h_n\in \C(\M(X),\C(X,\Rplus))$ with which the $H_i$ can be expressed: there is a recursive function
$\psi:\N^2\to\N$ such that for all $i$, $H_i=\sup_k h_{\psi(i,k)}$.

Furthermore, $h_n(\mu)$ (strictly speaking, $Eval(h_n,\mu)$, see proposition \ref{curry}) is bounded by a constant computable from $n$ and independent of $\mu$. Hence, the integration operator
$\int:\M(X)\times\N\to[0,1]$ which maps $(\mu,\uple{i_1,\ldots,i_k})$ to
$\int \sup\{h_{i_1}(\mu),\ldots,h_{i_k}(\mu)\}d\mu$ is computable.

We are now able to define $T$: $T(i,\mu)=\sup\{H^k_i(\mu):\int H^k_i(\mu)d\mu<1\}$ where $H^k_i=\sup_{n<k}h_{\psi(i,n)}$. As $\int H^k_i(\mu)d\mu$ can be computed from $i,k$ and a description of $\mu$, $T$ is a constructive function from $\N\times \M(X)$ to $\C(X,\Rplus)$.
\end{proof}

As a consequence, every randomness test for a particular
measure can be extended to a uniform test:

\begin{theorem}[Uniformity vs non-uniformity]\label{theorem_uniform}
Let $\mu_0$ be a measure. For every $\mu_0$-randomness test $t$ there is a
uniform randomness test $T:\M(X)\to\C(X,\Rplus)$ with $T(\mu_0)=\frac{1}{2}t$.
\end{theorem}

\begin{proof}
let $\mu_0$ be a measure and $t$ a $\mu_0$-randomness test: $\frac{1}{2}t$ is then
a $\mu_0$-constructive element of the enumerative lattice $\C(X,\Rplus)$, so by lemma
\ref{lemma_extensional} there is a constructive element $H$ of
$\C(\M(X),\C(X,\Rplus))$ such that $H(\mu_0)=\frac{1}{2}t$. There is some $i$
such that $H=H_i$: $T_i$ is a uniform randomness test satisfying
$T_i(\mu_0)=\frac{1}{2}t$ because $\int H_i(\mu_0)d\mu_0=\frac{1}{2}\int td\mu_0<1$.
\end{proof}

\begin{theorem}[Universal uniform test]\label{theorem_universal_test}
There is a universal uniform randomness test, that is a uniform test $T_u$ such that for every uniform test $T$ there is a constant $c_T$ with $T_u\geq c_T T$.
\end{theorem}

\begin{proof}
it is defined by $T_u:=\sum_i 2^{-i-1}T_i$: as every $T_i$ is a uniform
randomness test, $T_u$ is also a uniform randomness test, and if $T$
is a uniform impossibility test, then in particular $\frac{1}{2}T$ is a
constructive element of $\C(\M(X),\C(X,\Rplus))$, so
$\frac{1}{2}T=H_i$ for some $i$. As $\int H_i(\mu)d\mu=\frac{1}{2}\int
T(\mu)d\mu < 1$ for all $\mu$, $T_i(\mu)=H_i(\mu)=\frac{1}{2}T(\mu)$ for
all $\mu$, that is $T_i=\frac{1}{2}T$. So $T_u\geq 2^{-i-2}T$.
\end{proof}







\begin{definition}\label{random_any_measure}
Given a measure $\mu$, a point $x\in X$ is called
\defin{$\boldsymbol{\mu}$-random} if $T_u^\mu(x)<\infty$. Equivalently, $x$ is
$\mu$-random if it lies in no $\mu$-effective null set.
\end{definition}

The set of $\mu$-random points is denoted by $R_\mu$. This is the
complement of the maximal $\mu$-effective null set $\{x\in X:T_u^\mu(x)=+\infty\}$.

\subsection{Randomness on a computable probability space}

We study the particular case of a computable measure. As a morphism of computable probability spaces is compatible with measures and computability structures, it shall be compatible with algorithmic randomness. Indeed:

\begin{proposition}\label{morphism_random}
Morphisms of computable probability spaces are defined on random points and preserve randomness.
\end{proposition}

To prove it, we shall use the following lemma:

\begin{lemma}\label{random_re_full}
In a computable probability space $(\X,\mu)$, every random point lies in every r.e open set of full measure. 
\end{lemma}

\begin{proof} let $U=\bigcup_{\uple{i,j}\in E} B(s_i,q_j)$ be a r.e open set of
  measure one, with $E$ a r.e subset of $\N$. Let $F$ be the r.e set
  $\{\uple{i,k}:\exists j, \uple{i,j}\in E, q_k<q_j\}$. Define:
$$
U_n=\bigcup_{\uple{i,k}\in F\cap[0,n]} B(s_i,q_k) \quad \mbox{ and } \quad \comp{V}_n=\bigcup_{\uple{i,k}\in F\cap[0,n]} \overline{B}(s_i,q_k)
$$

Then $U_n$ and $V_n$ are r.e uniformly in $n$, $U_n\nearrow U$ and $\comp{U}=\bigcap_n V_n$. As $\mu(U_n)$ is
lower semi-computable uniformly in $n$, a sequence $(n_i)_{i\in\N}$
can be computed such that $\mu(U_{n_i})>1-2^{-i}$. Then
$\mu(V_{n_i})<2^{-i}$, and $\comp{U}=\bigcap_i V_{n_i}$ is a
$\mu$-Martin-L\"of test. Therefore, every $\mu$-random point is in $U$.
\end{proof}

\begin{proof}(of proposition \ref{morphism_random})
let $F:D\subseteq X\to Y$ be a morphism. From lemma \ref{random_re_full},
every random point is in $D$ which is an intersection of full-measure r.e
open sets.

Let $t:Y\to\Rplus$ be the universal $\nu$-test. The function $t\circ F:D\to\Rplus$ is lower
semi-computable. Let $\A$ be any algorithm lower semi-computing it: the associated lower
semi-computable function $f_\A:X\to\Rplus$ extends $t\circ F$ to the whole
space $X$ (see lemma \ref{lemma_extensional}). As $\mu(D)=1$, $\int
t\circ F d\mu$ is well defined and equals $\int f_\A
d\mu$. As $F$ is measure-preserving, $\int t\circ F d\mu=\int td\nu\leq
1$. Hence $f_\A$ is a $\mu$-test. Let $x\in X$ be a $\mu$-random point: as
$x\in D$, $t(F(x))=f_\A(x)<+\infty$, so $F(x)$ is $\nu$-random.
\end{proof}

\begin{corollary}
Let $(F,G):(\X,\mu)\rightleftarrows(\mathcal{Y},\nu)$ be an isomorphism of
computable probability spaces. Then $F_{|_{R_\mu}}$ and $\restr{G}{R_\nu}$ are total computable bijections
between $R_\mu$ and $R_\nu$, and $(\restr{F}{R_\mu})^{-1}=\restr{G}{R_\nu}$.
\end{corollary}

In particular:

\begin{corollary}\label{random_comp_measure}
Let $\delta$ be a binary representation on a computable probability space $(\X,\mu)$. Each point having a $\mu_\delta$-random expansion is $\mu$-random and each $\mu$-random
point has a unique expansion, which is $\mu_\delta$-random.
\end{corollary}

This proves that algorithmic randomness over a computable probability space
could have been defined encoding points into binary sequences using a
binary representation: this would have led to the same notion of
randomness. Using this principle, a notion of Kolmogorov complexity
characterizing Martin-L\"of randomness comes for free. For $x\in D$,
define:
$$
H_n(x)=H(\omega_{0..n-1}) \mbox{ and } \Gamma_n(x)=\delta([\omega_{0..n-1}])
$$
where $\omega$ is the expansion of $x$ and $H$ is the prefix Kolmogorov
complexity.

\begin{corollary}\label{random_complexity}
Let $\delta$ be a binary representation on a computable probability space $(\X,\mu)$. Then $x$ is $\mu$-random if and only if there is $c$ such that for all $n$:
$$
H_n(x)\geq -\log \mu(\Gamma_n(x)) - c
$$
\end{corollary}

All this allows to treat algorithmic randomness within probability theory over general metric spaces. In \cite{GalHoyRoj07b} for instance, it is applied to show that in ergodic systems over metric spaces, algorithmically random points are well-behaved: they are \textit{typical} with respect to any ergodic endomorphism of computable probability space, generalizing what has been proved in \cite{Vyu97} for the Cantor space.

\section*{Aknowledgments}
We would like to thank Stefano Galatolo, Peter G\'acs and Giuseppe Longo for useful comments and remarks.

\bibliography{bibliography}{}
\bibliographystyle{elsart-num}

\end{document}